# Data Mining Application to Attract Students in HEI

Umesh Kumar Pandey
Research scholar,
Singhania University Pacheri Bari Jhunjhunu, Rajasthan
umesh6326@gmail.com

Surjeet Kumar Yadav
Dept. of MCA,
VBS Purvanchal University, jaunpur
surjeet_k_yadav@yahoo.co.in

Saurabh pal
Head, Dept. of MCA,
VBS Purvanchal University,
Jaunpur, UP
drsaurabhpal@yahoo.co.in

*Abstract*—In the last two decades, number of Higher Education Institutions (HEI) grows in leaps and bounds. This causes a cut throat competition among these institutions while attracting the student get admission in these institutions. To make reach up to the students institution makes effort of advertisement. Similarly developing and developed both type of institution launch several services also to attract students. Most of the institutions are opened in self finance mode. So all time they feel short hand in expenditure. Now a day a number of advertisement methods are available. So it is difficult for an institution to make advertisement through all modes and launch all services at the same time due to different constraints. In this paper we use support and confidence method to find out the best way of advertisement.

## I. INTRODUCTION

Business world used data mining technique to find out the interest group for its product from a large number of consumers. No doubt they got satisfaction from data mining. This successful implication of data mining motivates researcher of education to use this technique because of rich data set. Actually data mining process consist mainly in discovery some valid, new, possibly useful and comprehensible structures from the dates [5].

In last few years Indian Higher Education Authorities focuses on opening self fund generating institution. In early days, these institutions easily attract students, who are seeking for admission; sometimes without any making special effort to attract them. But increased number of self finance HEI has faced a lot of trouble in attracting student. To solve this problem HEI's started thinking about some method to attract student.

One of the different solutions is marketing of the institution like business world which focuses on the market and shows its products quality to the consumer. In the case of HEI, students are like a customer who come and join the institution. In this context a new field is emerging known as Strategic Enrollment Management.

Association rule are frequently used by retail stores to assist marketing, advertising, floor placement and predicting faults in telecommunication networks. These uncovered relationships are not inherent in the data as with functional dependencies and they don't represent any sort of causality or correlation [4].

In this paper we try to find the best advertising method to market a Higher education Institution.

## II. ENROLLMENT MANAGEMENT

The term enrollment management was first used in a recruiting and retention context in the early 1970 by John "Jack" Maguire at Boston.[6]. Strategic planning provides a clear direction for an institution and allows its leaders to chart the proper course. It gives you the courage to challenge.

Self finance institution always has a fear of economic uncertainty. Strategic enrollment management was used not to address only admission but financial aids, retention, and marketing and with more students looking at institution composition as a selling point, campus demographics.

Competing on Analytics, Davenport and Harris identified a typology of query, reporting, and analytics capabilities that organizations can use to improve performance and competitive position. Figure 1 illustrates how these capabilities can be deployed to support strategic enrollment management.

## Application to Strategic Enrollment Management (SEM)

| Type of Analytics or Reporting | Focus | Application to Strategy |
| --- | --- | --- |
| **Analytics** | | |
| Optimization | What's the best that can happen? | Refine SEM strategies, targets, practices, and actions to achieve optimal outcomes. |
| Predictive Modeling | What will happen next? | Sophisticated predictive modeling shapes recruitment, yield, and retention policies and practices. |
| Forecasting/ Extrapolation | What if these trends continue? | Forecasting SEM strategies and outcomes. |
| Statistical Analysis | Why is this happening? | Current and longitudinal analysis to support refinement and execution of SEM strategy. |
| **Query & Reporting** | | |
| Alerts | What actions are needed? | Alerts and interventions, based on both predictive modeling and actual current performance and level of engagement. |
| Query/Drill Down | Where exactly is the problem? | Capacity to drill down to examine individual learners based on dynamic viewing of cohorts. |
| Ad Hoc Reports | How many, how often, where? | Capacity to expand list of variables and dynamically view cohorts and individual learners using many variables. |
| Standard Reports | What happened? | Standard portfolio of reports and views most needed by decision makers |

Adapted from Davenport and Harris 2007

Figure 1: [http://www.semworks.net/papers/wp_Metrics-and-Analytics-in-SEM.php]

Many institutions are using combinations of predictive modeling and statistical analysis to recruit, improve the performance of their admissions funnel, and design programs and policies to improve retention. [1, 2, 6, 13]

### III. DATA MINING

Data mining, knowledge discovery and machine learning are used in same context. They contains algorithm to find best pattern from an unstructured data with the help of computers. These algorithms attempt to fit a model to the data. The algorithms examine the data and determine a model that is closest to the characteristics of the data being examined [4].

Data mining or knowledge discovery in databases (KDD) is the automatic extraction of implicit and interesting patterns from large data collections. KDD can be used not only to learn the model for the learning process or student modeling but also to evaluate and to improve e-learning systems by discovering useful learning information from learning portfolios [12]

### IV. RELATED WORK

Data mining provides a great aid to education by finding hidden pattern from unstructured student database. A lot of research is done in educational data mining area using data mining. Yadav and Pal [14, 15] make a comprehensive study on different research did in the area of education data mining. They also provide a list of data mining techniques.

Mcintyre C [7] discussed three phased work which should provide breakthrough for colleges and universities that struggle with enrollment forecasting and enrollment management by effectively integrating both sets of activities. Commonly available and frequently used, measures of institutional and student: performance, including recruitment and retention, are key features of this work.

Yadav, Bharadwaj and Pal [12] obtained the university students data like attendance, class test, seminar and assignment marks from the students' database, to predict the performance at the end of the semester using three algorithms ID3, C4.5 and CART and shows that CART is the best algorithm for classification of data.

Bohannon T [3] discussed various application of predictive modeling in higher education. His paper concentrates on enrollment management, retention analysis and donor giving. He used a decision tree, backward regression, stepwise regression and a neural network data mining technique to complete his study.

Pandey and Pal [11] used association rule to find the student interest of choosing class language. In this paper they use seven different interestingness measure to find the students interest. They concluded that student has shown their interest in mix mode class language.

Merceron and Yacef [8] were interested in detecting association mistakes done by student using interestingness measures such as lift, conviction, correlation etc. They explored the interestingness measures under different variant of the data sets.

Bharadwaj and Pal [17], applied the classification as data mining technique to evaluate student' performance, they used decision tree method for classification. The goal of their study is to extract knowledge that describes students' performance in end semester examination. They used students' data from the student' previous database including Attendance, Class test, Seminar and Assignment marks. This study helps earlier in identifying the dropouts and students who need special attention and allow the teacher to provide appropriate advising.

V. APPLICATION

In this study data gathered from a degree college named PSRIET, Ranjeetpur Chilbila Pratapgarh, UP, affiliated with Dr. RMLA University, Faizabad, India. College has completed its 2 year having total strength of around 350 students of streams BA, BCA and PGDCA. College has decided to make advertisement to attract more students. In this sequence college collected data from, already admitted, student by asking question - "How s/he knows about the college first time?"

*A. Data collection:*

A college is situated in the municipal area of Pratapgarh. It is not a high-tech city. So data is first collected on enquiry paper, then all data stored into the computer to make analysis. In this enquiry form one section is more important which provide data for this analysis i.e. advertisement method by which s/he knows about the college

*B. Data Analysis:*

In last two year college uses traditional method of advertisement to attract student because most of the population is not using modern method of advertisement like internet etc. Data collected is shown in the following table:

TABLE I: NUMBER OF OCCURRENCE OF ADVERTISEMENT METHODS

| Advertisement method | Code | Answer |
|---|---|---|
| Hording | H | 230 |
| News paper | N | 160 |
| Pamphlets | P | 100 |
| Radio | R | 30 |
| Advertisement Van | V | 50 |
| Personal Contact | C | 130 |

In every enquiry form two advertisements has been mentioned. In the next table we have shown how many number of times different combination occur while answering the advertisement method.

TABLE II: OCCURRENCES OF DIFFERENT COMBINATION OF ADVERTISEMENT

| Relation | Occurrence |
|---|---|
| Hording and Newspaper (HN) | 120 |
| Hording and Pamphlets (HP) | 70 |
| Hording and Radio | 10 |
| Hording and Per. contact(HC) | 30 |
| Newspaper and Pamphlets (NP) | 20 |
| Newspaper and P. Contact(NC} | 20 |
| Pamphlets and P. Contact(PC) | 10 |
| Radio and P. Contact (RC) | 20 |
| Advt. Van and P. Contact | 50 |

*C. Support and confidence Analysis:*

The support of an item (or a set of items) is the percentage of transactions in which that item occurs. It is defined as "the support (s) for an association rule X→Y is the percentage of transactions in the database that contain X∪Y.

The confidence or strength ($\alpha$) for an association rule X→Y is the ratio of the number of transactions that contain X∪Y to the number of transactions that contain X.

Confidence measures the strength of the rule whereas support measures how often it should occur in the database. Typically, large confidence values and a smaller support are used. Support and confidence value of the collected data is shown in following table:

TABLE III : SUPPORT AND CONFIDENCE OF DIFFERENT RELATION

| Relation | S | $\alpha$ |
|---|---|---|
| N→H | 0.3429 | 0.7500 |
| P→H | 0.2000 | 0.7000 |
| R→H | 0.0286 | 0.3333 |
| P→N | 0.0571 | 0.2000 |
| C→H | 0.0857 | 0.2308 |
| N→C | 0.0571 | 0.4571 |
| P→C | 0.0286 | 0.2857 |
| C→R | 0.0571 | 0.6667 |
| C→V | 0.1429 | 0.3714 |

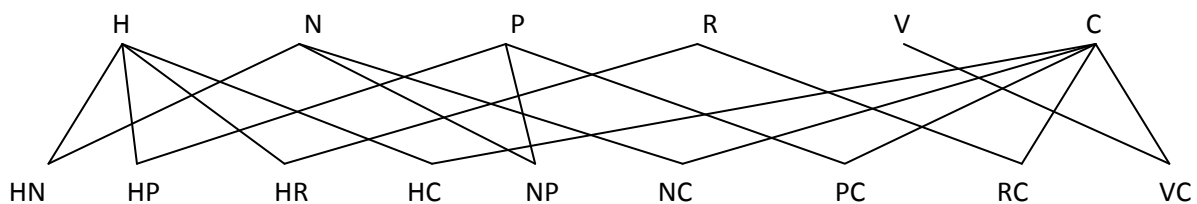

Figure 2: relation with different existing advertisement method in data

Umesh Kumar Pandey et al. / International Journal on Computer Science and Engineering (IJCSE)

*D.  Cosine analysis:*

Cosine is a number between 0 and 1. A value close to 1 indicates a good correlation between x and y i.e. the more transaction containing x also contains y and vice versa.. The closer cosine to 0 the more transactions contain x without containing y and vice versa. Cosine value is calculated as below: [8]

$$Cosine(X \Rightarrow Y) = \frac{P(X,Y)}{\sqrt{P(X)*P(Y)}}$$

Cosine value for the collected data is shown in following table:

TABLE IV: COSINE VALUE OF DIFFERENT RELATION

| Relation | Cosine value |
|---|---|
| N→H | 0.6255 |
| P→H | 0.4616 |
| R→H | 0.1204 |
| P→N | 0.1581 |
| C→H | 0.1735 |
| N→C | 0.1387 |
| P→C | 0.0877 |
| C→R | 0.3203 |
| C→V | 0.6202 |

*E.  Result:*

From table III we can conclude that most of the student around 34.29% have answered newspaper and hording as a best medium of advertisement; and the confidence level shows that 75% of the time newspaper occurs so does hording. Similarly support and confidence level are showing for other relations. From table III and figure 2 we can say that the entire advertisement mediums are connected to either hording or personal contact. In figure 2 hording has four edges and personal contact has five edges.

Another result is obtained from table IV by cosine analysis. It shows that newspaper → hording and personal contact and advertisement van has good relation but other are showing that they are not closely related.

## VI.  CONCLUSION AND FUTURE WORK

In this paper we tried to find out the best method of advertisement. In table 1 most of the students have opted hording (230 times) and newspaper (160 times). But after performing data mining techniques on this data we concluded that the hording and personal contact could be good method for attracting students because both of them are associated with all other method of advertisement.

In future we will use cost factor associated with each advertisement. This will help an institution to make an economic and effective advertisement method.

AUTHORS PROFILE

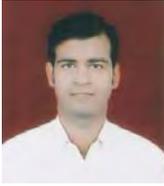
Umesh Kumar Pandey is Assistant Professor in the Department of Computer Applications, P S R I E T, Pratapgarh, UP India. He obtained his M.C.A degree from IGNOU (2004) and M.Phil. in Computer Science from PRIST University, Tamilnadu. He is currently doing research in Data Mining and Knowledge Discovery from Singhania University, Rajasthan. Umesh Kumar Pandey published two papers in international Journals i.e. IJCSI & IJCSIT and one paper in national journal i.e. LBSIMDS. He participated in one national conference in LBSIMDS. He is member of two international associations i.e. IAENG and IACSIT.

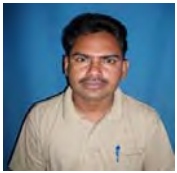
Surjeet Kumar Yadav received his M.Sc. (Computer Science) from Dr. Baba Sahed Marathwada University, Aurangabad, Maharastra, India (1998). At present, he is working as Sr. Lecturer at Department of Computer Applications, VBS Purvanchal Uniersity, Jaunpur. He is an active member of CSI and National Science Congress. He is currently doing research in Data Mining and Knowledge Discovery.

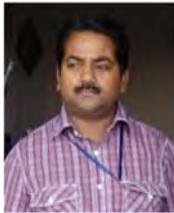
Saurabh Pal received his M.Sc. (Computer Science) from Allahabad University, UP, India (1996) and obtained his Ph.D. degree from the Dr. R. M. L. Awadh University, Faizabad (2002). He then joined the Dept. of Computer Applications, VBS Purvanchal University, Jaunpur as Lecturer. At present, he is working as Head and Sr. Lecturer at Department of Computer Applications.
Saurabh Pal has authored a commendable number of research papers in international/national Conference/journals and also guides research scholars in Computer Science/Applications. He is an active member of CSI, Society of Statistics and Computer Applications and working as reviewer and member of editorial board for more than 15 international journals. His research interests include Image Processing, Data Mining, Grid Computing and Artificial Intelligence.